\documentclass[
    ,final            
  ]
  {aipproc}

\layoutstyle{8x11double}

\newcommand{\rxdixhuit}{RX\,J1856.5\-$-$3754}
\newcommand{\rxzerosept}{RX\,J0720.4\-$-$3125}
\newcommand{\rxseize}{RX\,\-J16\-05.3\-+3249}

\newcommand{\rxa}{RX\,J0806.4\-$-$4123}
\newcommand{\rxb}{RX\,\-J0420.0\-$-$5022}
\newcommand{\degr}{\ensuremath{^\circ}}
\begin{document}

\title{Proper motions of ROSAT discovered isolated neutron stars measured with Chandra: 
First X-ray measurement of the large proper motion of RX J1308.6+2127/RBS\,1223 }


\author{C. Motch}{address={CNRS, Observatoire de Strasbourg, Strasbourg, France }}

\author{A. M. Pires}{address={Instituto de Astronomia, Geof\'isica e Ci\^encias Atmosf\'ericas da USP, S\~ao Paulo, Brazil},altaddress={CNRS, Observatoire de Strasbourg, Strasbourg, France }}

\author{F. Haberl}{address={Max-Planck-Institut f\"ur extraterrestrische Physik, Garching, Germany}}

\author{A. Schwope}{address={Astrophysikalisches Institut Potsdam, Potsdam, Germany}}

\author{V.E. Zavlin}{address={Space Science Laboratory, NASA MSFC, Huntsville, AL, USA}}

\begin{abstract}
 The unprecedented spatial resolution of the Chandra observatory opens the possibility to detect with relatively high accuracy proper motions at X-ray wavelengths. We have conducted an astrometric study of three of the "Magnificent Seven", the thermally emitting radio quiet isolated neutron stars (INSs) discovered by ROSAT. These three INSs (RX J0420.0-5022, RX J0806.4-4123 and RX J1308.6+2127/RBS\,1223) either lack an optical counterpart or have one too faint to be used for astrometric purposes. We obtained ACIS observations 3 to 5 years apart to constrain or measure the displacement of the sources on the X-ray sky using as reference the background of extragalactic or remote galactic X-ray sources.

Upper limits of 138 mas/yr and 76 mas/yr on the proper motion of RX J0420.0-5022 and RX J0806.4-4123, respectively, have already been presented in \cite{motch2007}. Here we report the very significant measurement ($\sim$ 10 sigma) of the proper motion of the third INS of our program, RX J1308.6+2127/RBS\,1223. Comparing observations obtained in 2002 and 2007 reveals a displacement of 1.1 arcsec implying a yearly proper motion of 223 mas, the second fastest measured for the ROSAT discovered INSs. The source is rapidly moving away from the galactic plane at a speed which precludes any significant accretion of matter from the interstellar medium. Its transverse velocity of $\sim$ 740 (d/700\,pc)\,km/s might be the largest of the "Magnificent Seven" and among the fastest recorded for neutron stars. RX J1308.6+2127/RBS\,1223 is thus a young high velocity cooling neutron star. The source may have its origin in the closest part of the Scutum OB2 association about 0.8\,Myr ago, an age consistent with that expected from cooling curves, but significantly younger than inferred from pulse timing measurements (1.5\,Myr).
\end{abstract}


\maketitle


\begin{figure}[ht]
  \includegraphics[height=.25\textheight]{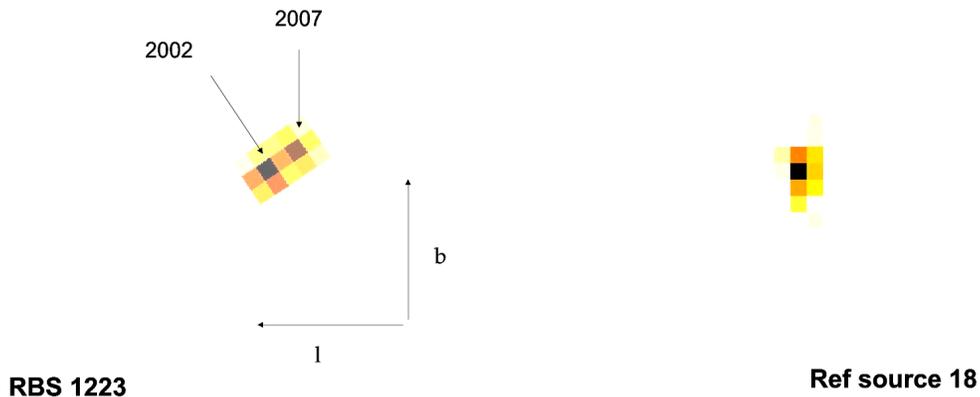}
  \caption{Chandra ACIS-I merged 2002 + 2007 images of RBS\,1223 and reference source 18. The position of the neutron star has moved by more than 2 ACIS-I pixels (1.1") after correction for the small residual shifts present between the 2002 and 2007 images. In contrast, the 2002 and 2007 images of reference source 18 fall on the same pixels.}
\end{figure}

\section{X-ray and optical properties of RX J1308.6+2127/RBS\,1223}

RBS 1223 is the fifth X-ray brightest of the "Magnificent Seven", the group of radio quiet and thermally X-ray emitting isolated neutron stars discovered by ROSAT. RBS\,1223 has average properties for the group, with kT = 86\,eV and a rotation period of 10.31\,s. A rather strong deviation from a purely Planckian energy distribution is observed below 500\,eV. This feature can be explained as a proton cyclotron absorption line in a field of 2 to 6 10$^{13}$\,G \cite{haberl2003}. Recent deep XMM-Newton observations point to a complex absorption 
structure consisting of two lines at 0.23\,keV and 0.46\,keV suggesting B $\sim$ 4 10$^{13}$\,G \cite{schwope2007}. However, an interpretation in terms of a pair of harmonics seems inconsistent with the ratio of the fluxes in the lines and alternative interpretation in terms of two distinct poles or atomic transitions may also be considered. A coherent timing solution linking Chandra and XMM-Newton pulse phase measurements allowed the determination of the slowing down of the period (Pdot = 1.12 10$^{-13}$\,s/s) which, if due to magnetic dipole torques predicts B = 3.4 10$^{13}$\,G, consistent with that estimated from the low energy absorption lines \cite{kaplan2005}. The age of 1.5\,Myr inferred from the slowing down of the rotation period is however much longer than the expected cooling time of a few 10$^{5}$ yr. Deep HST images show the presence of a very faint V $\sim$ 28 object in the Chandra error circle which is the likely counterpart of the X-ray source \cite{kaplan2002a}.

\section{Why are proper motions important?}
The kick imparted to the neutron star during its birth in the supernova explosion can accelerate the star to very large speeds, even larger than the Galactic escape velocity. A small fraction of the old INSs could however have low enough velocities to be re-heated by accretion of matter from the interstellar medium. The proper motion is thus a sensitive diagnostics of the X-ray powering mechanism. For young cooling objects, the motion vector provides hints on the birth place and thus age estimates which can then be compared with those expected from cooling curves assuming various mechanisms. 

\section{Detecting proper motions in X-rays}
For radio quiet neutron stars with no or too faint optical counterparts, X-rays constitute the only chance to measure their displacement on the sky. Unfortunately, in spite of motions of up to a few 100 mas/yr, the spatial resolution of most X-ray instruments remains too bad to detect the proper motion in a reasonable time interval. Few attempts to measure displacements in X-rays have been reported. An upper limit of $\sim$ 170 mas/yr on the motion of AXP 1E2259+586 is derived by \cite{ot2005} using ROSAT, Chandra and XMM-Newton data. Chandra HRC observations \cite{hui2006} yield a $\sim$ 3 $\sigma$ detection of the motion of Puppis-A. Finally, \cite{neu2001} detects a proper motion of 0.34 $\pm$ 0.12" for RX J1856.5-3754 in ROSAT HRI images, consistent with that formerly determined in the optical. 

Chandra ACIS I and S imagers offer the best ever long term stability and spatial resolution. We took advantage of these unprecedented performances to organize pairs of observations separated by 3 or 5 years using the same instrument setting, aimpoints, roll angle and time of the year. The exposure time and detector type were selected as to acquire a large enough number of galactic or extragalactic reference sources while avoiding pile-up on the main target. 

All data were reprocessed using the latest calibrations available and most recent version of CIAO (CIAO 3.4 and CIAO 4.0beta1). Source detection was performed using wavdetect testing different sensitivity thresholds and energy ranges on both randomized and de-randomized images. Using a maximum likelihood method, we computed from the set of sources common to both observations (typically 12 to 25) a transformation from one sky coordinate system to the other including rotation around the aimpoint and translation in right ascension and declination. Our MARX simulations show that relative motions can be retrieved by ACIS with an accuracy better than 0.1". A more detailed description of the method used and reports on the upper limits obtained for RX J0420.0-5022 and RX J0806.4-4123 are given in \cite{motch2007}.

\section{The proper motion of RBS\,1223}
The source was observed with ACIS-I in May 2002 and May 2007. A minimum error on the relative positions of the 2002 and 2007 astrometric frames of 68 mas is obtained using non randomized data, a 0.5-5\,keV band and a threshold of 10$^{-7}$ and a total of 12 common sources. The residual offsets between the 2002 and 2007 astrometric frames are 70 mas in right ascension and 260 mas in declination. Our results are however quite insensitive to the source detection parameters used and to the number of common reference sources considered.

Fig. 1 shows that the position of RBS\,1223 has clearly moved by 1100 mas between 2002 and 2007, corresponding to a total proper motion of 223 +/- 26 mas/yr. In contrast, the position of, for instance, source 18 remains unchanged. Taking into account all positioning errors (relative frame, RBS\,1223 and a systematic error of 70 mas \cite{motch2007}) yields :


$\mu_{\alpha}$ cos($\delta$) =  -207 $\pm$ 20 mas/yr

$\mu_{\delta}$   =      84 $\pm$ 20 mas/yr


With a proper motion of $\sim$ +185 mas/yr in galactic latitude, RBS\,1223 is rapidly escaping from the galactic plane, as expected for an object born from a massive star and now located at $b$ = 83.08\degr.

The proper motion of RBS\,1223 is thus the second largest of the "Magnificent Seven" (see Table1). The distance to the source is not constrained by the interstellar absorption \cite{posselt2007}. Assuming blackbody emission and similar radii for all ROSAT INSs yields d$\,\sim\,$670\,pc \cite{kaplan2002b}, whereas the light curve modeling of \cite{schwope2005} argues in favor of shorter distances in the range of 76 to 380\,pc. A transverse velocity of $\sim$ 740 (d/700\,pc)\,km/s precludes any significant accretion of matter from the interstellar medium. At a distance of 670\,pc, RBS\,1223 has the fastest transverse velocity observed for nearby neutron stars. Overall, the velocity distribution of ROSAT INSs does not appear significantly different from that of radio pulsars (see Fig. 2).

\begin{figure}[ht]
  \includegraphics[height=.3\textheight]{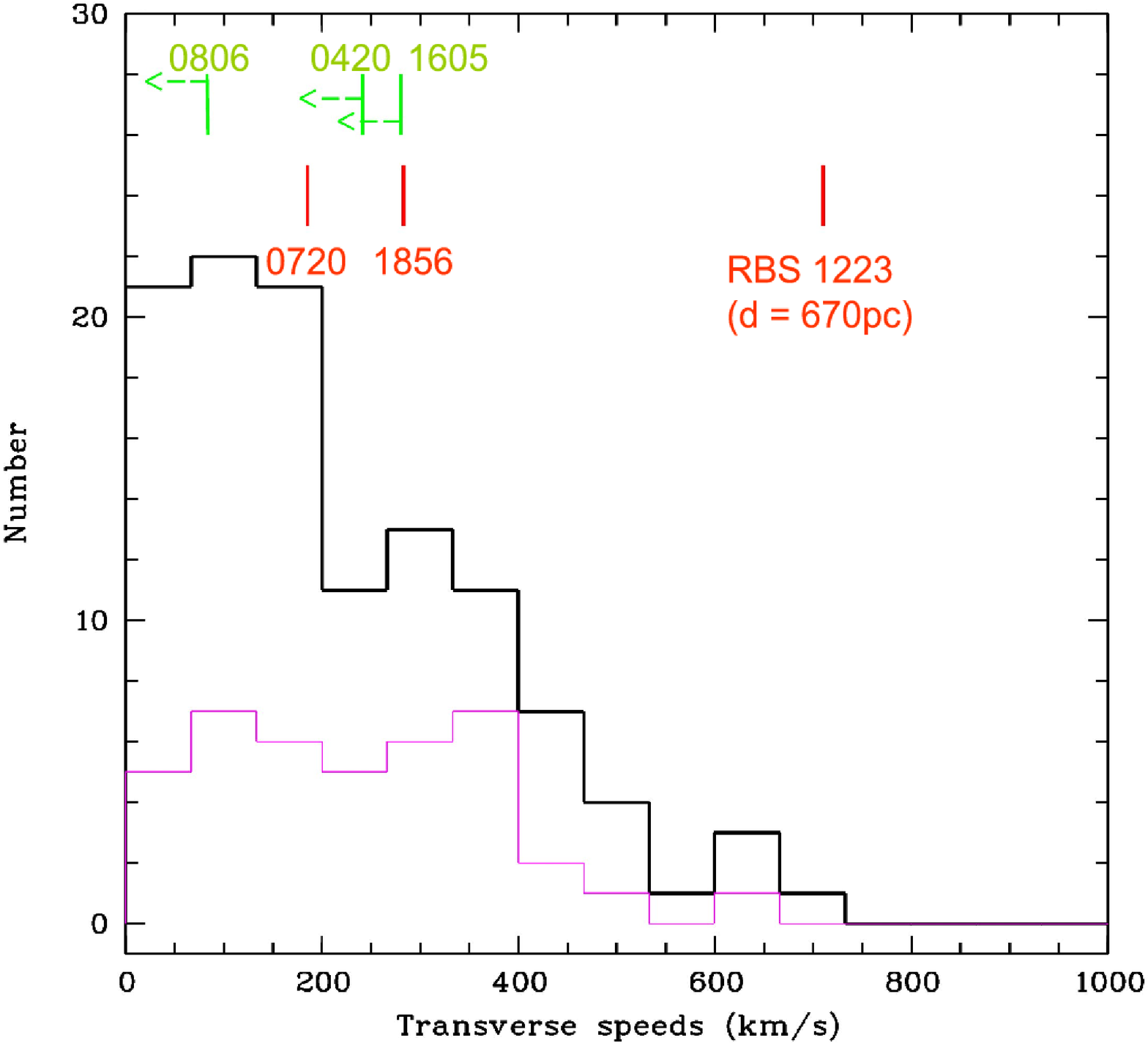}
  \caption{Transverse velocities of the ROSAT discovered INSs compared to those of nonrecycled radio pulsars. Data taken from \cite{hobbs2005}. The (black) thick line and the light (magenta) line show the histogram of the entire population and of the young population (age $<$ 3\,Myr) respectively.}
\end{figure}

\section{Possible origin and age}

Although the accuracy of the motion vector derived here cannot rival that provided by ground based or HST observations, Chandra data can nevertheless efficiently constrain the range of possible birth places and flight times. We find that most possible backward trajectories of RBS\,1223 intersect the galactic plane away from catalogued OB associations. However, a relatively small fraction of them cross the Scutum OB2 A and B associations at different ages. Basically, two groups of solutions exist. First, a "young" age of 0.8\,Myr with a birth place in the closest part (d=510\,pc) of the OB association, a receding velocity of $\sim$ 600\,km/s and a present distance of $\sim$ 500\,pc. An age of 0.8\,Myr is consistent with the expected cooling time but is at variance with that derived from pulse timing. Second, an "older" age of 1.2\,Myr, more consistent with the observed spin down, with a birth place in the closest or the farthest part (d = 1170\,pc) of Scutum OB2, receding velocities of 300 to 600\,km/s and present distances of 400 to 800\,pc.  Very few trajectories pass at relatively short distances from the Upper Sco part of the nearby OB2 association about 0.4 to 0.8\,Myr ago. Although these "nearby" solutions are much less likely, the inferred present distance of 175\,pc is then consistent with that derived by Schwope et al. (2005). We conclude that if one rules out ages older than 1\,Myr on the basis of its inconsistency with cooling curves, the most likely birth place of RBS\,1223 is in Scutum OB2 A about 0.8\,Myr ago, suggesting that the age derived from spin down is probably biased as already noted by \cite{kaplan2005}.


\begin{table}
\label{propermotions}
\begin{tabular}{llll}
\noalign{\smallskip}
Name            &   Prop. mot.   &  distance  	& V$_{\rm T}$  \\ 
                &   (mas/yr) 	   &	(pc)		& (km\,s$^{-1}$) \\
\hline
\rxdixhuit      &   332$\pm$1	   & 178$^{+22}_{-17}$  & 283 \\ 
{\bf RBS\,1223}  & 223$\pm$26       & 76 - 670           & 80-710\\
\rxzerosept     &   97$\pm$12	   & 360$^{+170}_{-90}$	& 185 \\ 
\rxseize        & 144.5$\pm$13.2   &  $<$ 410		& $<$ 280\\ 
\rxb            & $\leq$138        & $\leq$340          & $<$ 241\\
\rxa            & $\leq$ 76        & 240$\pm$25         & $<$ 83\\
RBS 1774        &  -               & $\sim$ 400         & - \\
\hline
Geminga         & 170$\pm$4 	   & 273$\pm$84 	& 219\\ 
B1929+10        & 103.4$\pm$0.2    &361$^{+10}_{-8}$	& 176 \\ 
Vela Pulsar     & 58.0$\pm$0.1	   & 287$^{+19}_{-17}$  & 79\\ 
B0656+14        &  44.1$\pm$0.7    & 288$^{+33}_{-27}$  & 60.1 \\ 
\noalign{\smallskip}
\hline
\end{tabular}
\caption{Proper motions, distances and transverse velocities of the "Magnificent Seven" and of nearby young pulsars}
\end{table}



\bibliographystyle{aipproc}   


\bibliography{sample}



\end{document}